\documentclass[useAMS,usenatbib]{mnras}
 
\usepackage{amsmath}
\usepackage{float}
\usepackage{graphicx}
\usepackage{txfonts}
\usepackage{indentfirst}
\usepackage{color}
\usepackage{ulem}
\usepackage{hyperref}
\usepackage[Symbolsmallscale]{upgreek}
\usepackage[export]{adjustbox}
    
\newcommand{\eq}[1]{\begin{equation}  #1 \end{equation}}

\newcommand{\br}[1]{\left( #1 \right)}

\newcommand{\ba}[1]{\left\langle #1 \right\rangle}

\newcommand{\dd}{{\rm d}}
\newcommand{\expo}[1]{~{\rm e}^{ #1 }}
\newcommand{\vek}[1]{\mbox{\boldmath $#1$}}

\newcommand{\msun}{\mbox{$\,{\rm M}_{\sun}$}}

\def\apj{ApJ}
\def\apjl{ApJ}
\def\apjs{ApJS}
\def\aap{A\&A}

\def\mnras{MNRAS}
\def\nat{Nature}

\title[Turbulence decay in stratified ICM]{Turbulence decay in the
density-stratified intracluster medium}

\author[Xun Shi and Congyao Zhang]{Xun Shi$^{1,2}$\thanks{E-mail:
xun@ynu.edu.cn}, 
Congyao Zhang$^{2}$ \\
$^{1}$South-Western Institute for Astronomy Research (SWIFAR), Yunnan
University, 650500 Kunming, P. R. China\\ 
$^{2}$Max-Planck-Institut f\"ur Astrophysik (MPA),
Karl-Schwarzschild-Stra{\ss}e 1, D-85740 Garching bei M\"unchen, Germany
}

\begin{document}


\maketitle
  
\label{firstpage}
\begin{abstract}
Turbulence evolution in a density-stratified medium differs from that of
homogeneous isotropic turbulence described by the Kolmogorov picture. We
evaluate the degree of this effect in the intracluster medium (ICM) with
hydrodynamical simulations. 
We find that the buoyancy effect induced by ICM density stratification
introduces qualitative changes to the turbulence energy evolution, morphology,
and the density fluctuation - turbulence Mach number relation, and likely
explains the radial dependence of the ICM turbulence
amplitude as found previously in cosmological simulations.
A new channel of energy flow between the kinetic and the potential
energy is opened up by buoyancy.
When the gravitational potential is kept constant with time, this
energy flow leaves oscillations to the energy evolution, and leads to a balanced
state of the two energies where both asymptote to power-law time evolution with slopes
shallower than that for the turbulence kinetic energy of homogeneous isotropic
turbulence. We discuss that the energy evolution can differ more significantly
from that of homogeneous isotropic turbulence when there is
a time variation of the gravitational potential.
Morphologically, ICM turbulence can show a
layered vertical structure and large horizontal vortical eddies in the
central regions with the greatest density stratification. In addition, we find
that the coefficient in the linear density fluctuation - turbulence Mach number
relation caused by density stratification is in general a variable with
position and time.
\end{abstract}


\section[]{Introduction}
According to our current knowledge of cosmic structure formation, galaxy
clusters are the largest virialized objects on the top of the structure growth
hierarchy in our Universe today. They form via merging and accretion of smaller
units of matter, which would generate turbulent motions in their reservoir of
diffuse hot gas -- the intracluster medium (ICM), as demonstrated by hydrodynamic numerical simulations
\citep[e.g.][]{norman99,iapichino08, nelson12}. Some recent dedicated
simulations have investigated the statistical properties of the turbulent motions in the ICM, mainly their compressibility and spectral properties
 \citep[e.g.][]{min14, miniati15, vazza09,vazza17}. 

Observationally, the
ubiquitous of turbulence in the ICM is confirmed by a
direct detection of the non-thermal broadening of the X-ray emission lines by the \textsl{Hitomi}
satellite \citep{hitomi16}, as well as indirect observations of the magnetic field fluctuations in the diffuse cluster
radio sources \citep[][]{murgia04, vogt05, bonafede10,vacca10, vacca12}, X-ray
surface brightness fluctuations or pressure fluctuations inferred from X-ray maps and 
Sunyaev-Zel'dovich effect maps \citep{schuecker04, churazov12, 
walker15, khatri16, zhuravleva18}, and the suppression of resonant
line scattering in the X-ray spectra \citep{churazov04, zhuravleva13, hitomi18}.
In the near future, direct measurements of ICM turbulence will reach an
unprecedented sensitivity thanks to the planned \textsl{XRISM} and
\textsl{Athena} satellites (see \citealt{simionescu19} for a recent review).

The theoretical framework that is widely used for turbulence studies is the
idealized Kolmogorov picture of homogeneous and isotropic turbulence. However,
the ICM is neither homogeneous nor isotropic, it is in rough hydrostatic
equilibrium with the gravitational field that bounds it, and as a consequence, 
its density is radial-dependent i.e. stratified. If and how would this density
stratification influence ICM turbulence is the topic of this study. 

In previous numerical studies of ICM gas motions, there are hints of a radial
dependence that is neglected by the Kolmogorov picture. Studies of
cluster samples in cosmological hydrodynamical simulations have discovered
a slightly increasing amplitude of non-thermal ICM gas motions with cluster radius
despite that the thermal velocity a.k.a. sound speed decreases with radius
\citep{lau09, bat12, nelson14b, shi15}. In the analytical model of
the ICM non-thermal pressure by \citet{shi14}, a radial-dependent turbulence
dissipation time is responsible for this radial dependence. The success of this
analytical model \citep{shi15,shi16} motivated \citet{shi18} to verify this
ansatz by performing a detailed multi-scale analysis of ICM turbulence evolution
in cosmological hydrodynamical simulations, where they found clear evidence of
faster turbulence dissipation in the central regions compared to the outer
regions of the ICM after injection at a major merger. In this paper, we
continue to explore the physical mechanism underlying such radial
dependent turbulence decay by investigating the effect of ICM density
stratification. We will introduce the physics of turbulence evolution in a
density-stratified medium in Sect.\;\ref{sec:theory}, describe our simulation
design in Sect.\;\ref{sec:methods}, and present our results in
Sect.\;\ref{sec:results}. Then we will discuss some physics neglected in this
study in Sect.\;\ref{sec:discussion} and conclude in
Sect.\;\ref{sec:conclusion}.

\begin{figure}
\centering
    \includegraphics[width=.37\textwidth]{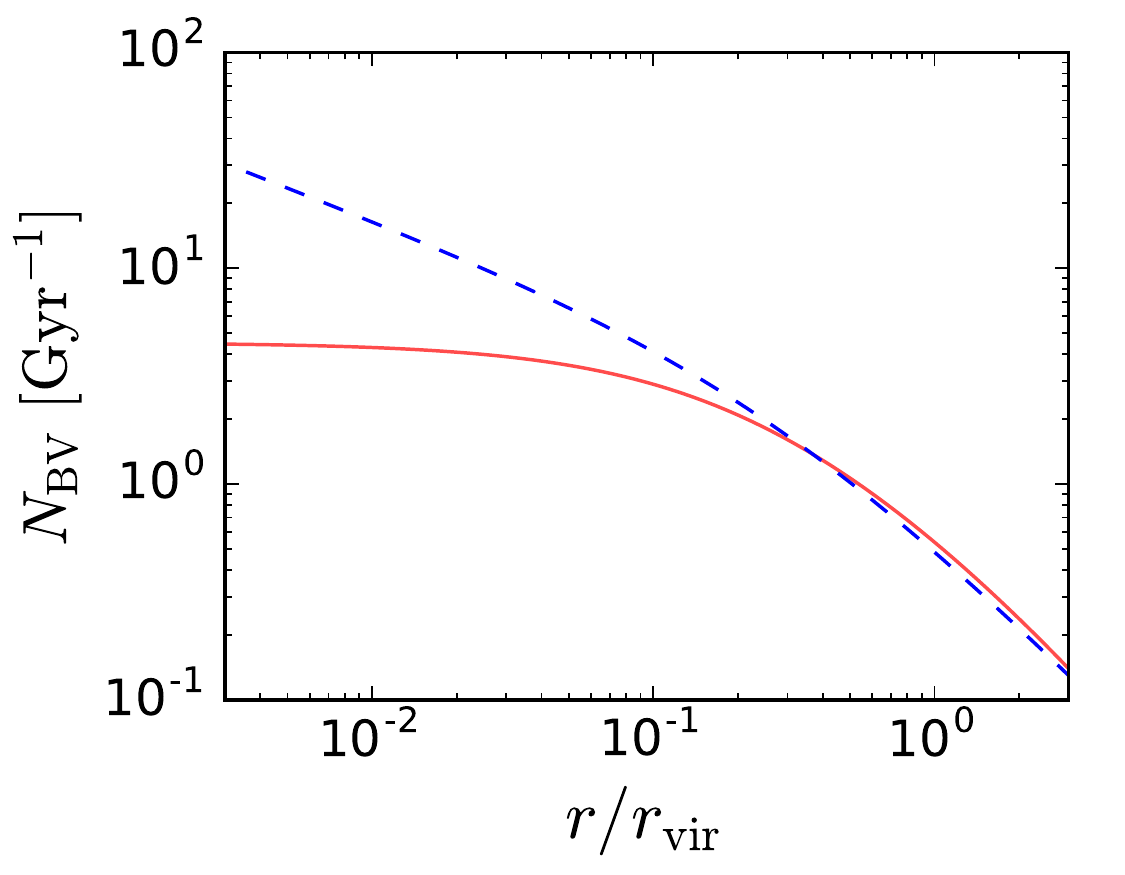} \\
     \includegraphics[width=.37\textwidth]{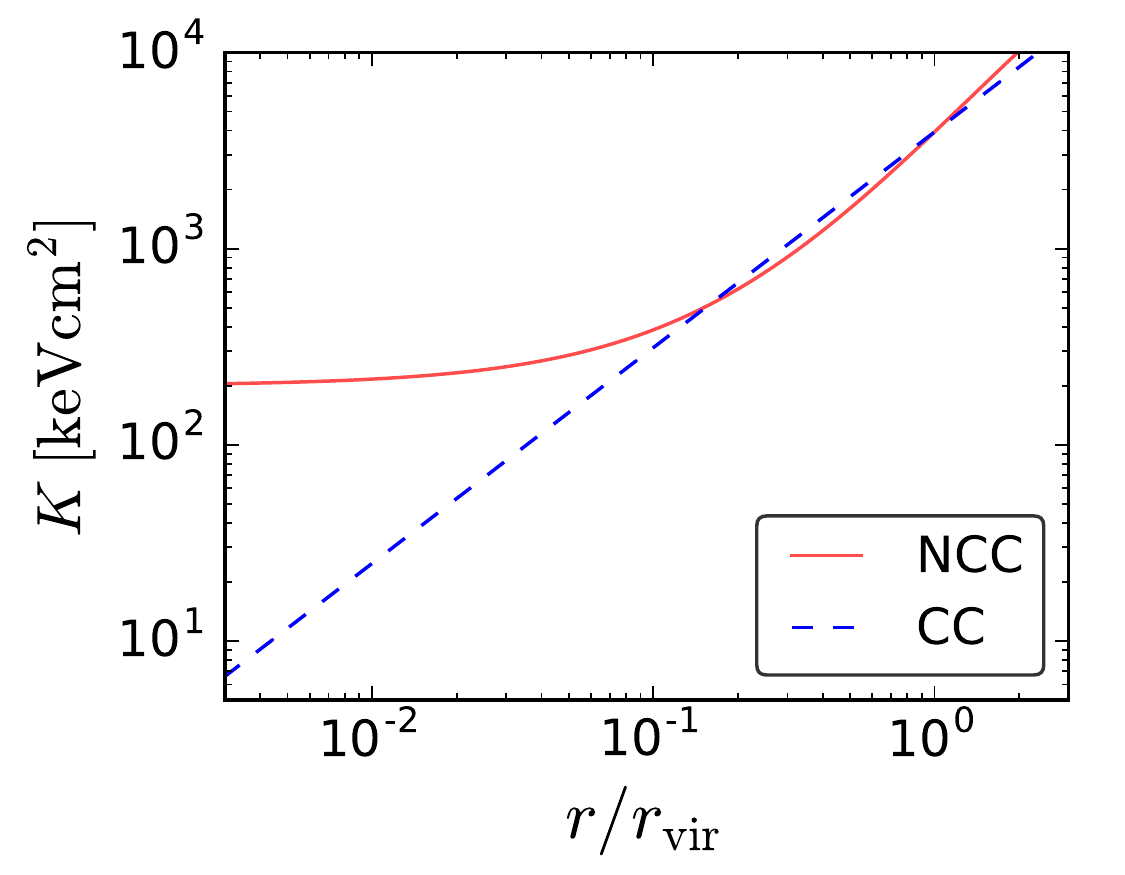} 
  \caption{Typical profiles of the Brunt-V\"ais\"al\"a
  frequency (Eq.\;\ref{eq:NBV}) in the ICM for cool-core (CC) and non-cool-core (NCC)
  clusters (top panel) and the corresponding entropy profiles (bottom panel). The NCC
  profiles are modelled with the \citet{komatsu01} analytical model for a
  $10^{15}\msun$ cluster at redshift zero. The CC profiles use the same
  NFW mass profile \citep{nfw96} as the NCC model, together with a typical
  entropy profile for a cool-core cluster, $K \propto r^{1.1}$.}
\label{fig:NBV}
\end{figure}

\section{Physics of Turbulence Evolution in a Density-Stratified Medium}
\label{sec:theory}
Turbulence evolution in a density-stratified system has been studied since long
in ocean and atmosphere sciences \citep[e.g.][]{ozmidov65, stillinger83, hopfinger87}. 

In a density-stratified system such as the ICM, gas motions lead inevitably to
density fluctuations. Also due to density stratification, the density fluctuations are
subject to a buoyancy force, which in turn alters the velocity field.
In other words, there exists a fundamental difference between turbulence in a
stably density-stratified medium and its homogeneous counterpart -- the presence
of restoring forces in the vertical direction. 

The buoyancy restoring force introduces a new time scale to the system, which is
characterized by the Brunt-Vais\"al\"a frequency
\eq{
\label{eq:NBV}
N_{\rm BV} = \sqrt{-\frac{g}{\gamma} \frac{\dd \ln K}{\dd r}} \,,
}
which is determined by the amplitude of the background gravitational
acceleration $\vek{g}$ and the entropy profile $K(r)$ of the galaxy cluster. Here $\gamma$ is the
adiabatic index of the ICM. As shown by Fig.\;\ref{fig:NBV},
the Brunt-Vais\"al\"a frequency differs by an order of magnitude from
the cluster center to the virial radius for a typical non-cool-core cluster,
and even more for a typical cool-core cluster which has a steep $N_{\rm BV}$
profile in its core region \footnote{Some clusters may have a very flat
entropy profile $\dd \ln K /\dd r \approx 0$ in the center and an associated
drop of the Brunt-Vais\"al\"a frequency. However, this is usually limited to a
very small radial range up to a few tens of kpc.}.

The important parameter describing the relative strength of the
buoyancy force compared to the inertia of the turbulence is the Froude number
\eq{
\label{eq:Fr}
Fr = \frac{\sigma}{N_{\rm BV} \ell_{\rm g}}
} 
where $\sigma$ is the turbulence velocity dispersion and 
\eq{
\ell_{\rm g} = \frac{\int k^{-1}E(k) \dd k }{\int E(k) \dd k }
}
is a lateral integral scale of the turbulence energy spectrum $E(k)$, with
$2\uppi \ell_{\rm g}$ characterizing the energy-containing scale\footnote{What
we define here is a Froude number for the whole turbulence velocity field, in contrast to the eddy
Froude number which is defined for individual turbulence eddies.}. The Froude
number describes the ratio of the buoyancy and the typical turbulence eddy
turn-over time scales.
 Generally speaking, $Fr < \mathcal{O}(1)$ (buoyancy time $\lesssim$ turbulence eddy
turn-over time) suggests
 that the buoyancy force becomes dynamically important, whereas $Fr \gg 1$
 (buoyancy time $\gg$ turbulence eddy turn-over time) suggests homogeneous
 isotropic turbulence unaffected by density stratification \citep[e.g.][]{riley03}.

Apart from $\ell_{\rm g}$, another fundamental turbulence length scale
associated with density stratification is the
buoyancy scale
\eq{
L_{\rm b} = \frac{2\uppi \sigma}{N_{\rm BV}}\,,
}
which characterizes the
thickness of the shear layers in stratified turbulence. One can see that the
Froude number also reflects the ratio of the two length scales $Fr \propto
L_{\rm b} / \ell_{\rm g}$.

Turbulence under the influence of an extreme density stratification ($Fr \ll 1$) is 
morphologically resemblant to 2D turbulence. It develops vertically thin and
horizontally extended ($L_{\rm b} \ll \ell_{\rm g}$) pancake-like structures
\citep[e.g.][]{brethouwer07}. Its energy dissipation is also significantly
different from that described by the Kolmogorov picture, as the vertical
turn-over of its energy-containing eddies is strongly suppressed by the density
stratification. In the ICM, a typical turbulence Froude number lies in the
intermediate regime of $Fr \sim \mathcal{O}(1)$ (\citealt{shi18}, see also
Fig.\;\ref{fig:Fr}).
How much would ICM turbulence evolution in this regime differ from that of homogeneous
isotropic turbulence requires dedicated studies.

\section{Methods}
\label{sec:methods}

\begin{figure}
\centering
    \includegraphics[width=.42\textwidth]{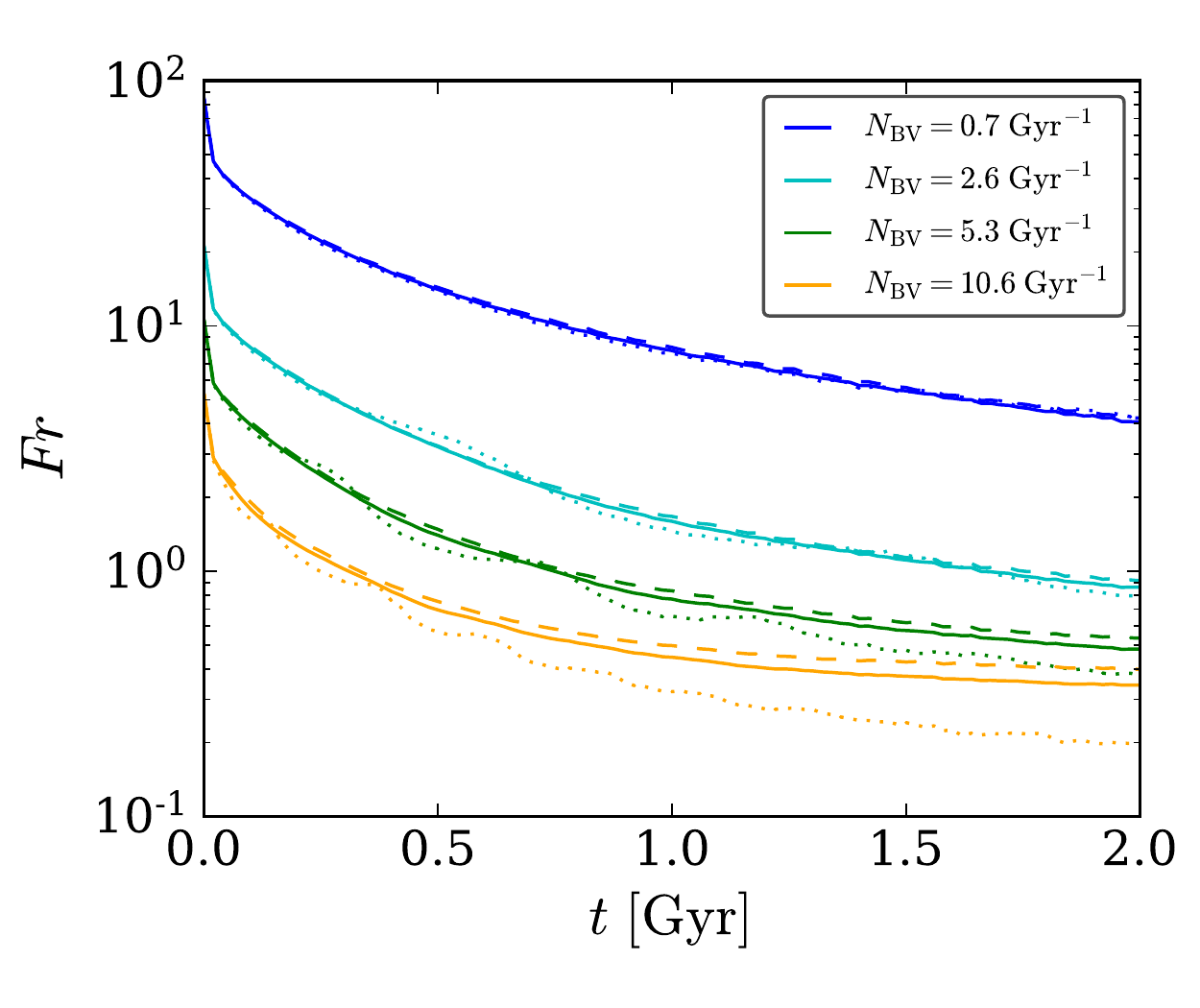} 
  \caption{Time evolution of the turbulence Froude number (solid lines). The
  dashed lines show the scaled horizontal Froude number defined using
  horizontal turbulence velocity and energy-containing scale, whereas the dotted
  lines show the corresponding scaled vertical Froude number.
  Their deviation from the solid lines reflects the velocity anisotropy.
  }
\label{fig:Fr}
\end{figure}

\begin{table*}
 \centering
{
\caption{Dimensionless parameters of the simulations. 
}
\begin{tabular}{cccccc}
\hline
 & Simulation runs & I & II & III & IV \\
  & $N_{\rm BV}$ [Gyr$^{-1}$] & 0.7 & 2.6 & 5.3 & 10.6 \\
\hline
$Fr_{\rm i}$  & Initial Froude number & $84.4$ &$21.1$   &$10.55$ &$5.27$ \\
$Fr_{\;\rm 2 Gyr}$ & Froude number at 2 Gyr & $4.1$ &$0.86$   &$0.48$ &$0.34$\\
$L_{\rm b, \;i} / L'_{z}$ & Initial buoyancy scale over vertical domain size
&
$12$ &$3.1$ &$1.5$ &$0.77$ \\
$L_{\rm b, \;2Gyr} / L'_{z}$ & Buoyancy scale at 2 Gyr over vertical domain size 
& $1.3$ &$0.35$ &$0.21$ & $0.13$ \\
$H_{\rho} / L'_{z}$ & Density scale height  over vertical domain size   
& $53$  &$13$   &$6.5$  &$3.2$ \\
\hline
\end{tabular}}
\label{tab:nondim_simparams}
\end{table*}

\begin{figure*}
\centering
\includegraphics[width=.8\textwidth]{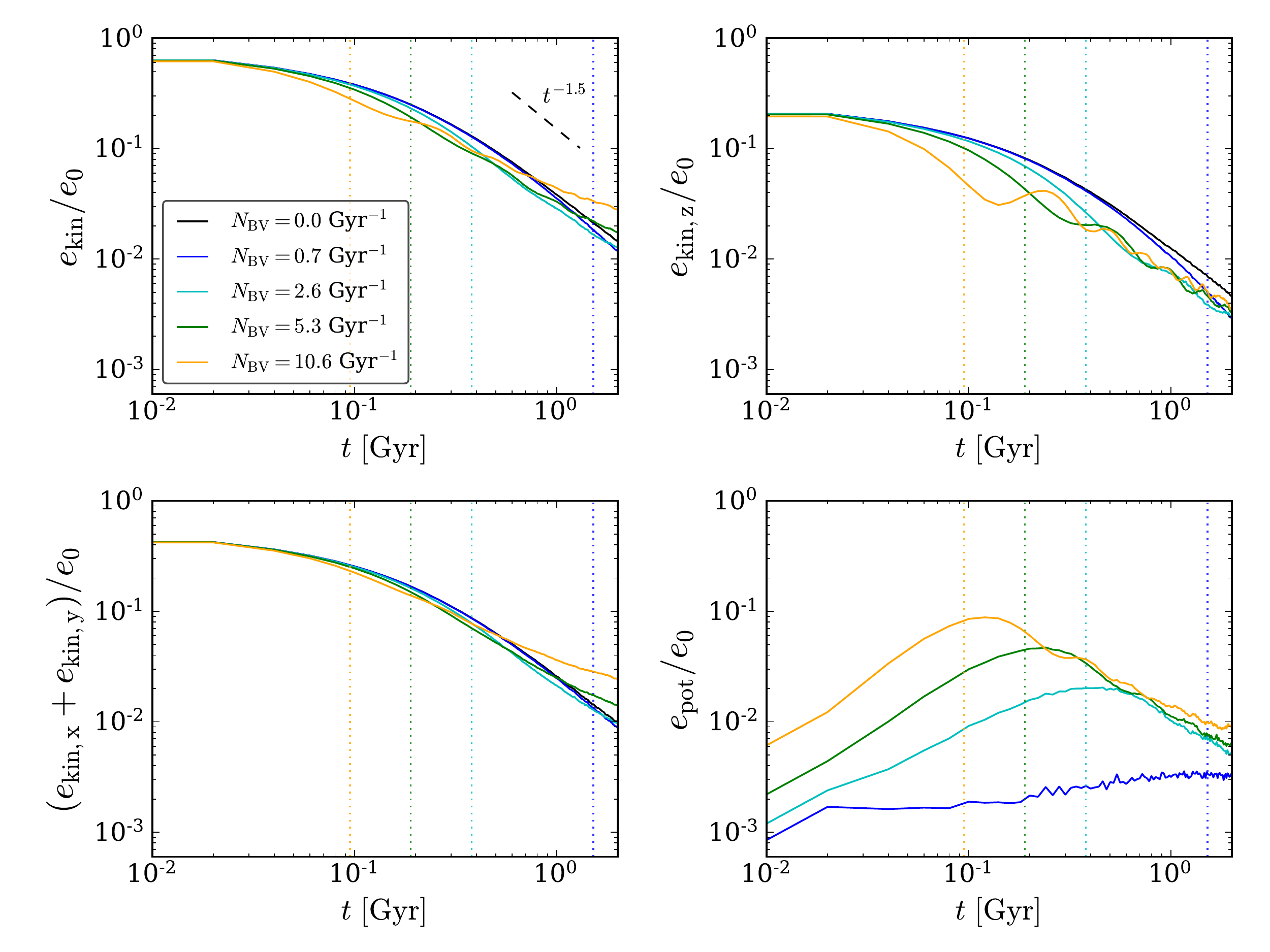}
\caption{Comparison of turbulence energy evolution in media with
different density stratification. Shown are the time evolution of the average
turbulence kinetic energy (top left panel), its vertical (top right panel)
and horizontal (bottom left panel) components, and the fluctuation gravitational
potential energy (bottom right panel) of the simulation runs with different
Brunt-Vais\"al\"a frequencies (colored lines) in comparison to those in a
homogeneous isotropic medium (black lines).}
\label{fig:compare_power_evo_low_mach}
\end{figure*}

\begin{figure}
\centering
\includegraphics[width=.4\textwidth]{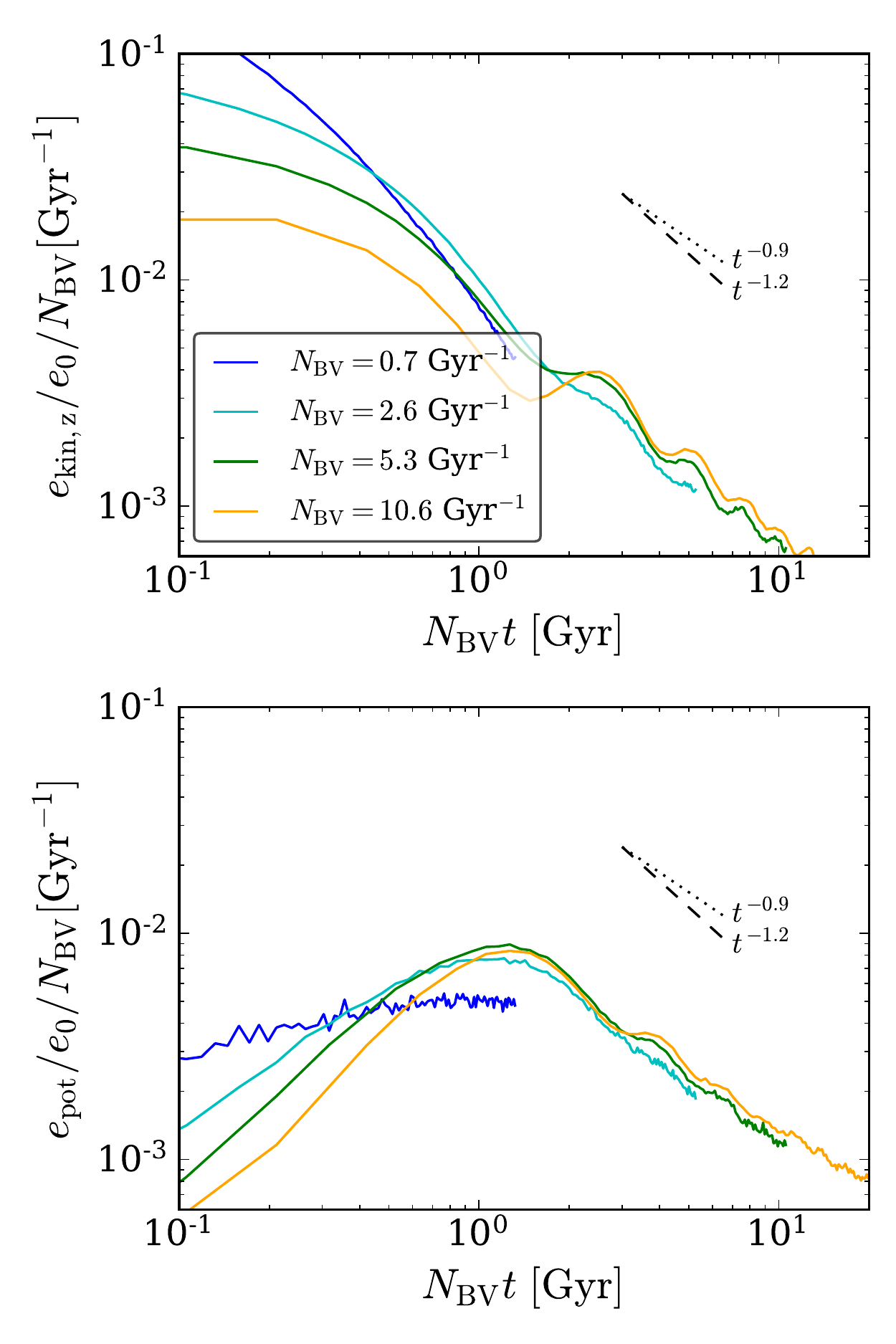}
\caption{Right panels of Fig.\;\ref{fig:compare_power_evo_low_mach} plotted with
scaled axes showing universal behavior at long times ($N_{\rm BV}t>1$).}
\label{fig:compare_power_evo_low_mach_NBVt_2panel}
\end{figure}

In this study, we perform idealized hydrodynamical numerical simulations to
examine the effect of density stratification on ICM turbulence evolution. 
We use the FLASH code \citet{fryxell00} to solve the hydrodynamics equations
\eq{
\frac{\partial \rho}{\partial t} + \nabla \cdot \br{\rho \vek{v}} = 0
}
and
\eq{
\label{eq:eom}
\frac{\partial \br{\rho \vek{v}}}{\partial t} + \nabla \cdot \br{\rho
\vek{v}\vek{v}} = -\nabla P + \rho \vek{g} }
for a polytropic gas in a stationary external gravitational field with a
constant gravitational acceleration $\vek{g}$ in the $-z$ direction. We simulate
local 3D boxes with $L_x = L_y = $ 200 kpc and $L_z = $ 250 kpc with a
resolution of $256 \times 256 \times 320$, and for a duration of $t_{\rm
max} = 2$ Gyr. 
Reflective hydrostatic boundary conditions are applied to top and bottom
boundaries, and periodic boundary conditions to the sides. 
We impose damping to the velocity fields near the top and bottom boundaries to
reduce boundary effects, and will omit the 20 grid points near the top and
bottom boundaries which are affected by the damping in the presentation below.
This leaves a vertical domain size of $L'_{z} = 218.75$ kpc.

The gas is chosen to have an exponential initial density profile $\rho_0 =
\bar{\rho} \expo{-z/H_{\rm \rho}}$ with a scale-height of 
\eq{
H_{\rm \rho} =  \frac{c_{\rm s}^2}{\gamma g}\,,
}
where the gas adiabatic index $\gamma = 5/3$ is relevant for fully
ionised ICM, and the sound speed $c_{\rm s}$ is chosen to be a constant across
the simulation domain. The corresponding initial
pressure
\eq{
P_0 =  \frac{c_{\rm s}^2 \rho_0}{\gamma}\,,
}
and $\rho_0$, $g$ satisfy the hydrostatic equilibrium condition. 

Our initial conditions are meant to mimic local ICM conditions in various
regions. We perform four simulation runs (I-IV) with different values of
gravitational acceleration $g$, corresponding to $N_{\rm BV} =$ 0.7
Gyr$^{-1}$, 2.6 Gyr$^{-1}$, 5.3 Gyr$^{-1}$ and 10.6 Gyr$^{-1}$, representing the
cluster outskirts, intermediate regions, central regions of non-cool-core
clusters, and central regions of cool-core clusters, respectively (Fig.\;\ref{fig:NBV}). 
We also include a simulation with no density stratification i.e.
$g=0$ as a comparison.

We use an identical initial turbulence velocity field for all simulation
runs to highlight the effect of density stratification on the subsequent
evolution. The initial velocity field is generated using a zero
mean Gaussian random field with an energy spectrum 
\eq{
E(k) \propto (k/k_{\rm c})^{-5/3}  \expo{-k_{\rm c}/k} \,,
}
mimicking a 3D turbulence spectrum with a Kolmogorov $k^{-5/3}$ inertial range
and a large scale cutoff $k_{\rm c} = 0.24$ kpc$^{-1}$. This initial energy
spectrum peaks at a length scale of $2\uppi\ell_{\rm g} = 43.6$ kpc. We
remove the compressive component to generate a solenoidal velocity field in correspondence
to the subsonic nature of typical intracluster turbulence. Then we normalize the
initial turbulence energy spectrum to have a velocity dispersion of $\sigma_0 =
278$ km/s and an initial specific kinetic energy $e_0 = \sigma_0^2 / 2$.
The resulting velocity field has several desired features e.g. statistically
homogeneity and isotropy, the proper energy spectrum and the complex,
three-dimensional structure expected of a turbulent flow field. 

The physical evolution of such an initial velocity field is fully dictated by
two dimensionless control parameters: the Froude number $Fr$ which reflects the
relative importance of the buoyancy force, and the Mach number $\mathcal{M}$
which shows the compressibility of the gas and its ability to generate sound
waves. In this study, we want to single out the buoyancy effect and thus
work in the low Mach number limit by choosing an unphysical high background
temperature $T = 512$ keV to suppress the generation of sound waves.  
We focus on the range of Froude number from $Fr \lesssim 1$ to $Fr \sim 10$
 relevant to galaxy clusters. The Froude number evolution of our four density-stratified
simulations are presented in Fig.\;\ref{fig:Fr}. We summarise the Froude
number and the other domain-size related dimensionless parameters of our
simulation runs in Table.\;1.

\begin{figure}
\centering
    \includegraphics[width=0.38\textwidth]{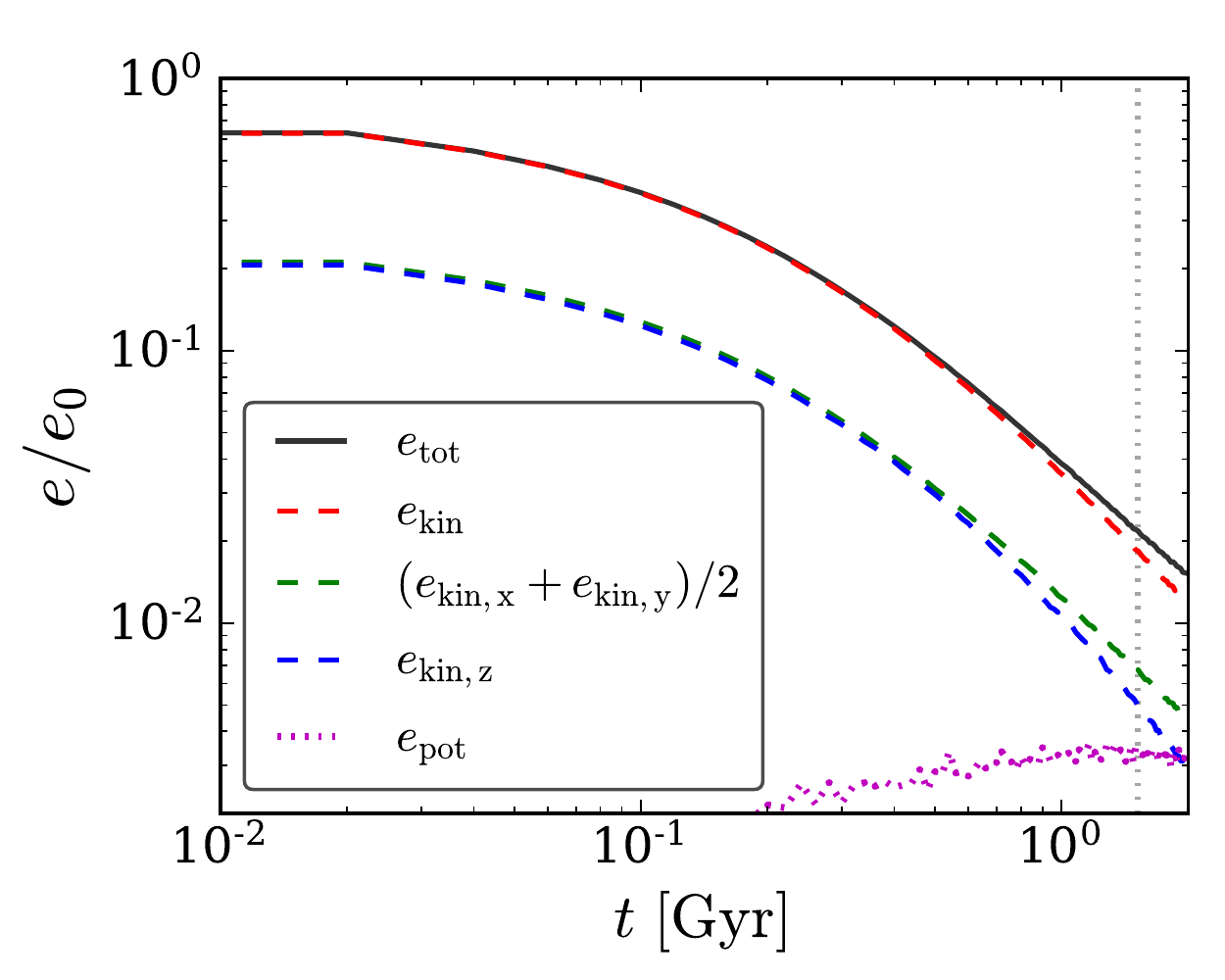} \\
    \includegraphics[width=0.38\textwidth]{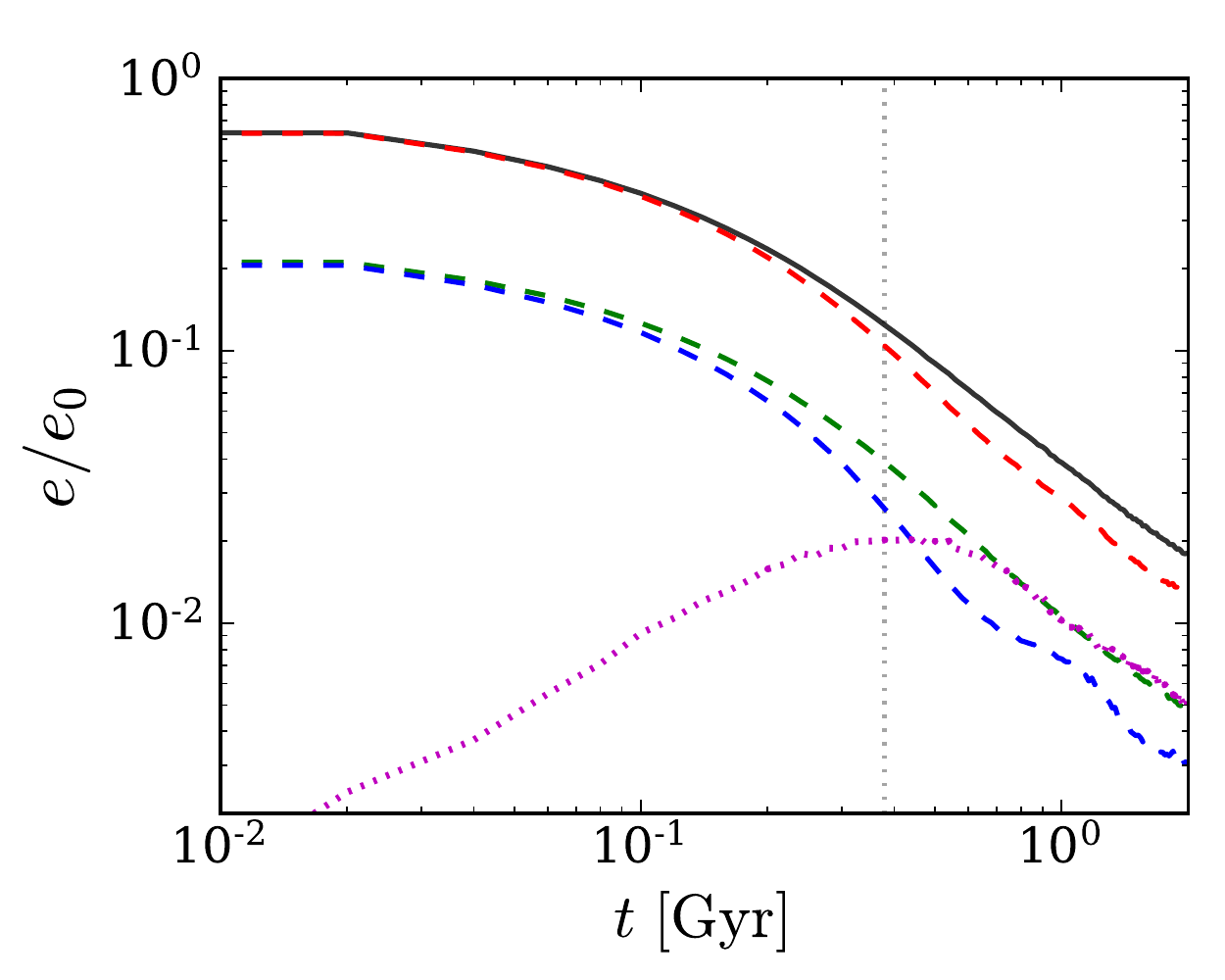} \\
    \includegraphics[width=0.38\textwidth]{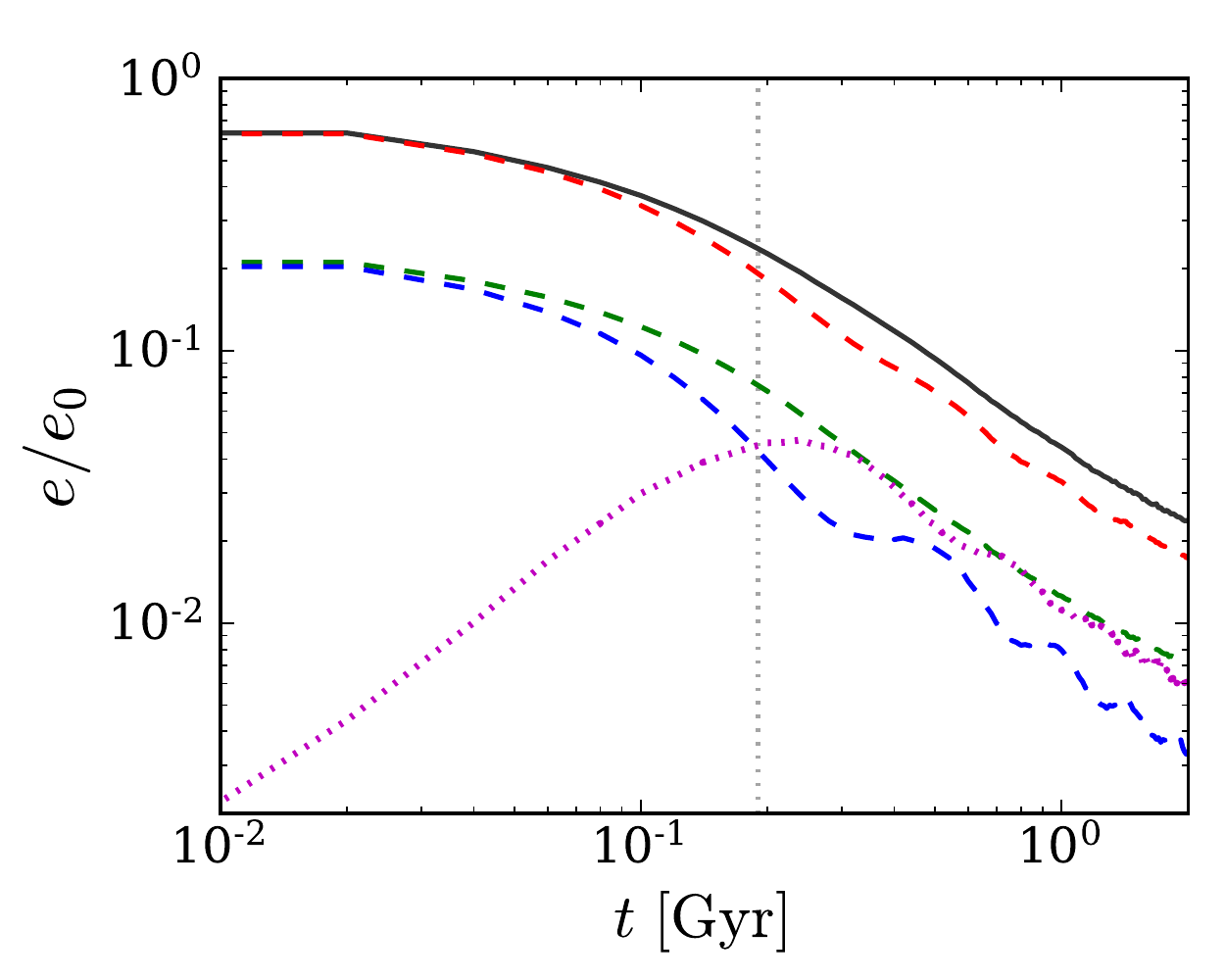} \\
    \includegraphics[width=0.38\textwidth]{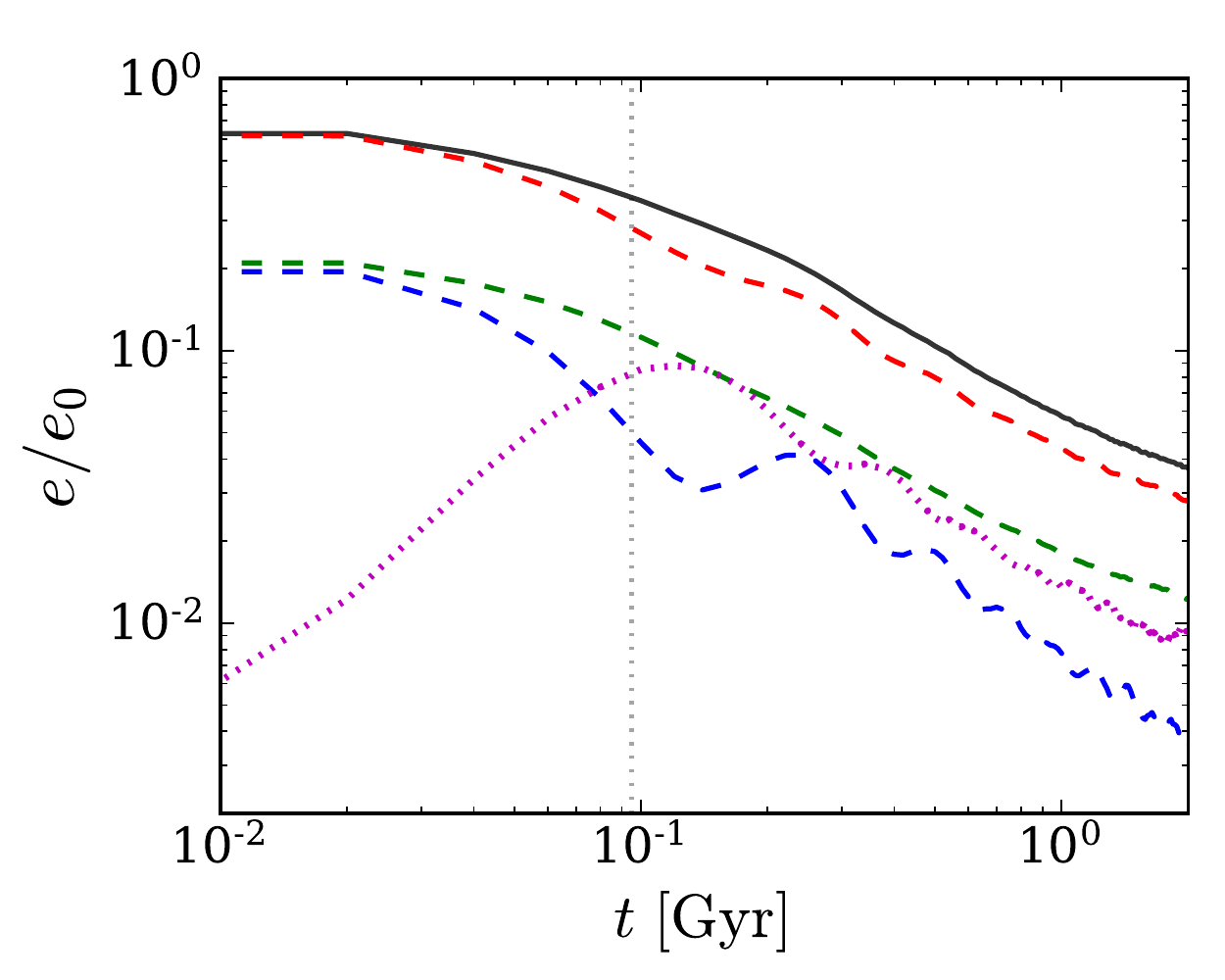} 
    \caption{Evolution of the energy densities for simulations with different
    density stratification (different panels). The vertical dashed line in each
    panel marks the Brunt-V\"ais\"al\"a time $1/N_{\rm BV}$ in the corresponding
    simulation. The red dashed line and the magenta dotted line mark the
    time evolution of turbulence kinetic energy and the fluctuation potential
energy respectively, and the black line is the sum of the two. The green/ blue
dashed lines represent the 1D horizontal/ vertical turbulence kinetic
energy.}
\label{fig:power_evo}
\end{figure}

\begin{figure*}
\centering
    \includegraphics[width=0.8\textwidth]{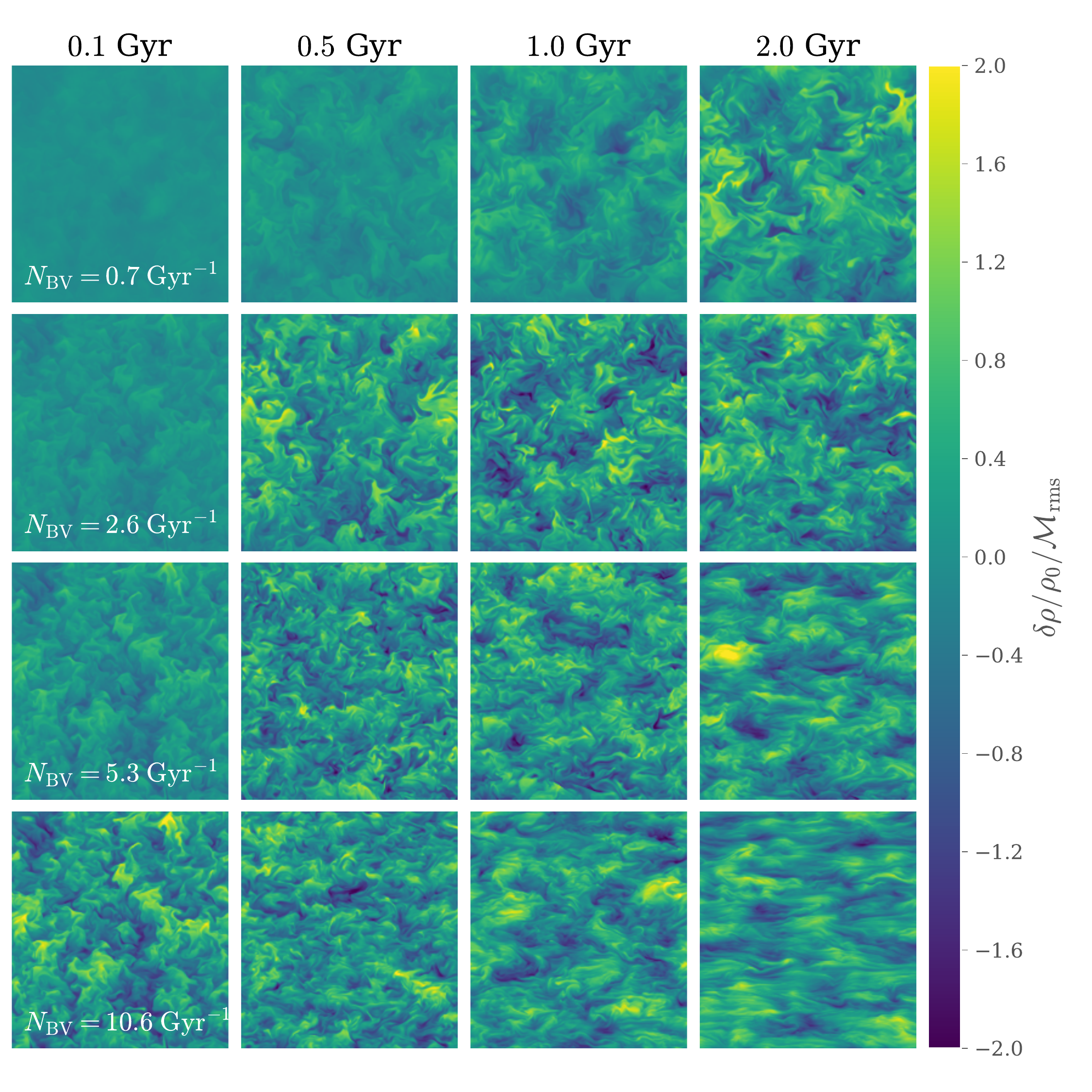}
\caption{Vertical $200$ kpc $\times$ $218.75$ kpc slices of density fluctuation
evolution through the mid-plane of the
x-axis for the simulation runs I-IV with various density gradients (rows) at four various times (columns) after turbulence injection at $t=0$.}
\label{fig:rho_yz}
\end{figure*}

\begin{figure*}
\centering
    \includegraphics[width=0.8\textwidth]{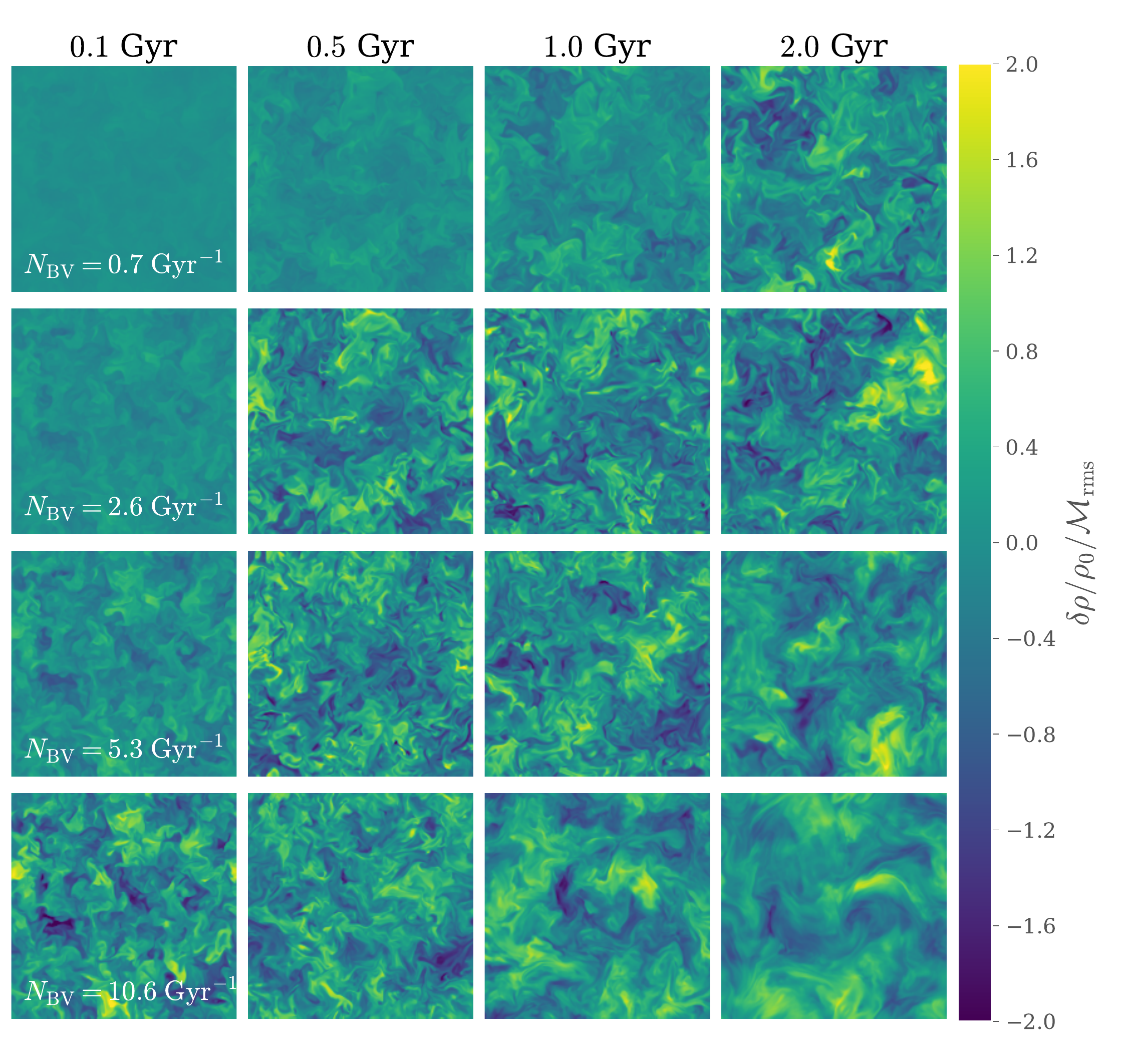}
    \caption{Same as Fig.\;\ref{fig:rho_yz}, but for horizontal $200$ kpc
    $\times$ $200$ kpc slices through the mid-plane of the z-axis. }
\label{fig:rho_xy}
\end{figure*}

\section{Results}
\label{sec:results}

\subsection{Energy evolution}
\label{sec:energy_evo}
\subsubsection{Kinetic-potential energy conversion and power law asymptotes}
For homogeneous, isotropic turbulence, the turbulence kinetic energy
decays in a power law fashion
\eq{
e_{\rm kin}(t) = \frac{\sigma^2}{2} \propto t^{-n} 
}
with an exponent $n \approx 1.5$ \citep[e.g.][]{landau59, frisch95,
sub06}.

Fig.\;\ref{fig:compare_power_evo_low_mach} shows a comparison of the evolution
of energy densities in our different simulation runs. Apparently, the turbulence
kinetic energy evolution (top left panel) in a density-stratified medium
(colored lines) differs from that in a homogeneous medium (black line) and shows much richer behavior. 
There are oscillations in the kinetic energy evolution in a density-stratified
medium, and the logarithmic slope of the kinetic energy time evolution asymptotes
to a value shallower than of $n\approx -1.5$ for homogeneous turbulence.
In the case of a high density stratification with $N_{\rm BV} = 10.6$ Gyr$^{-1}$, 
the logarithmic slope becomes as shallow as $-0.7$.

The oscillations in the kinetic energy evolution is created by a conversion
between the vertical kinetic energy (top right panel)
and the
fluctuation potential energy (bottom right panel) 
\eq{
e_{\rm pot} = - \frac{1}{2} \frac{g}{\rho_0} \br{\frac{\dd \rho}{\dd z}}^{-1}
\overline{\delta\rho^2}  \,,
}
as a result of the buoyancy induced by density stratification. Here,
$\delta \rho = \rho - \rho_0$ is the density fluctuation. As shown by
Fig.\;\ref{fig:power_evo}, this energy conversion leaves both the vertical
kinetic energy (blue dashed line) and the potential energy (magenta dotted
line) oscillate at opposite phases.

The time scale of this energy conversion is well captured by the typical
buoyancy time $1/N_{\rm BV}$ (vertical dotted lines). As shown by
Fig.\;\ref{fig:compare_power_evo_low_mach_NBVt_2panel}, the peak of the
fluctuation potential energy occurs at a time $t \approx 1.2/N_{\rm BV}$, and
both the vertical kinetic energy and the fluctuation potential energy asymptote
to a power-law evolution after the first period of the oscillation at $t
\gtrsim 1/N_{\rm BV}$. The long term asymptotes show roughly
universal shapes:
\eq{
e_{\rm kin, z}/e_{0} \propto N_{\rm BV} (N_{\rm BV} t)^{-1.2}
}
and
\eq{
e_{\rm pot}/e_{0} \propto N_{\rm BV} (N_{\rm BV} t)^{-0.9}\,,
}
if we neglect the low stratification run which has low signal to noise. This 
shows that (1) the long term decay of the vertical kinetic energy and the
fluctuation potential energy is governed by how many buoyancy periods the system
has evolved; (2) the power-law slopes are different, and both shallower than
$-1.5$; (3) their asymptotic amplitudes at the same time $t$ depends
little on $N_{\rm BV}$, and thus the amplitude at the same buoyancy period
$N_{\rm BV}t$ are roughly proportional to $N_{\rm BV}$; and (4) the amplitude of
the potential energy is greater than that of the vertical kinetic energy (see also
Fig.\;\ref{fig:power_evo} for a comparison of energy amplitudes for individual simulations). 

Another consequence of this vertical energy conversion is the emergence of
velocity anisotropy. 
The more stratified the background density is, the more anisotropic the
turbulence kinetic energy becomes, as reflected by the greater separation
between the vertical and horizontal turbulence kinetic energies
(Fig.\;\ref{fig:power_evo}). At long times, it is the horizontal component (bottom left
panel of Fig.\;\ref{fig:compare_power_evo_low_mach}) that contributes
more to the total kinetic energy, since density stratification suppresses 
turbulence eddy turn-over in the vertical direction especially for the large
energy-containing eddies. Consequently, the kinetic
energy of these large horizontal eddies can hardly cascade to
smaller scales and get dissipated, similar to the case in 2D turbulence. The
same suppression effect also results in the dominance of large eddies in the
horizontal plane, which will be shown in Sect.\;\ref{sec:morphology}.

\subsubsection{Radial dependence of ICM turbulence}
Let simulations with different density stratification represent different regions of the
ICM, the dependence of turbulence energy evolution on density stratification
then implies a radial-dependence of ICM turbulence. 

As shown by Fig.\;\ref{fig:compare_power_evo_low_mach}, in the first few hundred million
years after the injection of turbulence, the turbulence kinetic energy is
smaller where the density stratification is higher (top left panel),
because a greater part of its vertical component (top right panel) is converted into fluctuation
potential energy (bottom right panel).
However, at later times turbulence kinetic energy is actually larger 
where the density stratification is higher, as shown by the reversed
order of the colored lines in the top left panel at $t \gtrsim 0.5$ Gyr.
This is the result of two factors.
First, in the vertical direction, some of the energy stored as fluctuation
potential energy flows back to kinetic energy, and at long times the vertical
kinetic energy is rather independent of the density stratification (upper
right panel).
Second, turbulence evolving in a more density-stratified medium manages to keep
more horizontal kinetic energy at long times (bottom left
panel). 

This time reversal of the turbulence radial dependence is not found in
the \citet{shi18} detailed analysis of ICM turbulence decay in two galaxy
clusters selected from a set of cosmological numerical simulations. In both clusters studied by \citet{shi18},
a faster turbulence decay at small radii following injection by major mergers
leads to an increasing turbulence amplitude with cluster radius, and this
positive radial dependence stays even after a few Gyrs' evolution. Statistical
results of cosmological numerical simulations \citep[e.g.][]{lau09, bat12,
nelson14b} also support a positive radial dependence of turbulence being the
generic state of ICM. This apparent discrepancy is likely explained by the
neglect of the background density and gravitational potential variations in
the current study, as will be discussed in Sect.\;\ref{sec:discussion}.

\subsection{Morphology}
\label{sec:morphology}
Figs.\;\ref{fig:rho_yz} and \ref{fig:rho_xy} show the turbulence morphological
evolution as traced by the density fluctuation for the four simulation runs
representing different regions of the ICM. Two signs of density-stratification
influence are clear in the central regions of galaxy clusters
(simulation runs with $N_{\rm BV} = 5.3$ and 10.6 Gyr$^{-1}$) at late times: the loss of statistical isotropy
in the vertical plane (Fig.\;\ref{fig:rho_yz}), and the dominance of large
eddies in the horizontal plane (Fig.\;\ref{fig:rho_xy}). 

The degree of density-stratification influence is well-described by the buoyancy
period $N_{\rm BV} t$ or the Froude number (Fig.\;\ref{fig:Fr}), with $N_{\rm
BV} t \gtrsim 5$ and $Fr \lesssim 1$ marking the regime where the
influence on morphology is apparent. In an extreme case in the ICM where a
cool-core is left quiescent for $t \approx 2$ Gyr i.e.
since redshift $ \approx 0.2$ (bottom right panel), the Froude number
reaches as low as $Fr \approx 0.3$, and a clear layered structure with thickness $L_{\rm b} <
\ell_{\rm g}$ develops. The velocity anisotropy and density fluctuation
associated with this layered structure are in principle observable from X-ray
line profile and surface brightness fluctuation, albeit challenging due to their
small amplitude.

\subsection{Density fluctuation - Mach number relation in the stratified ICM}
\label{sec:rhov_scatter}

\begin{figure}
\centering
\includegraphics[width=.47\textwidth]{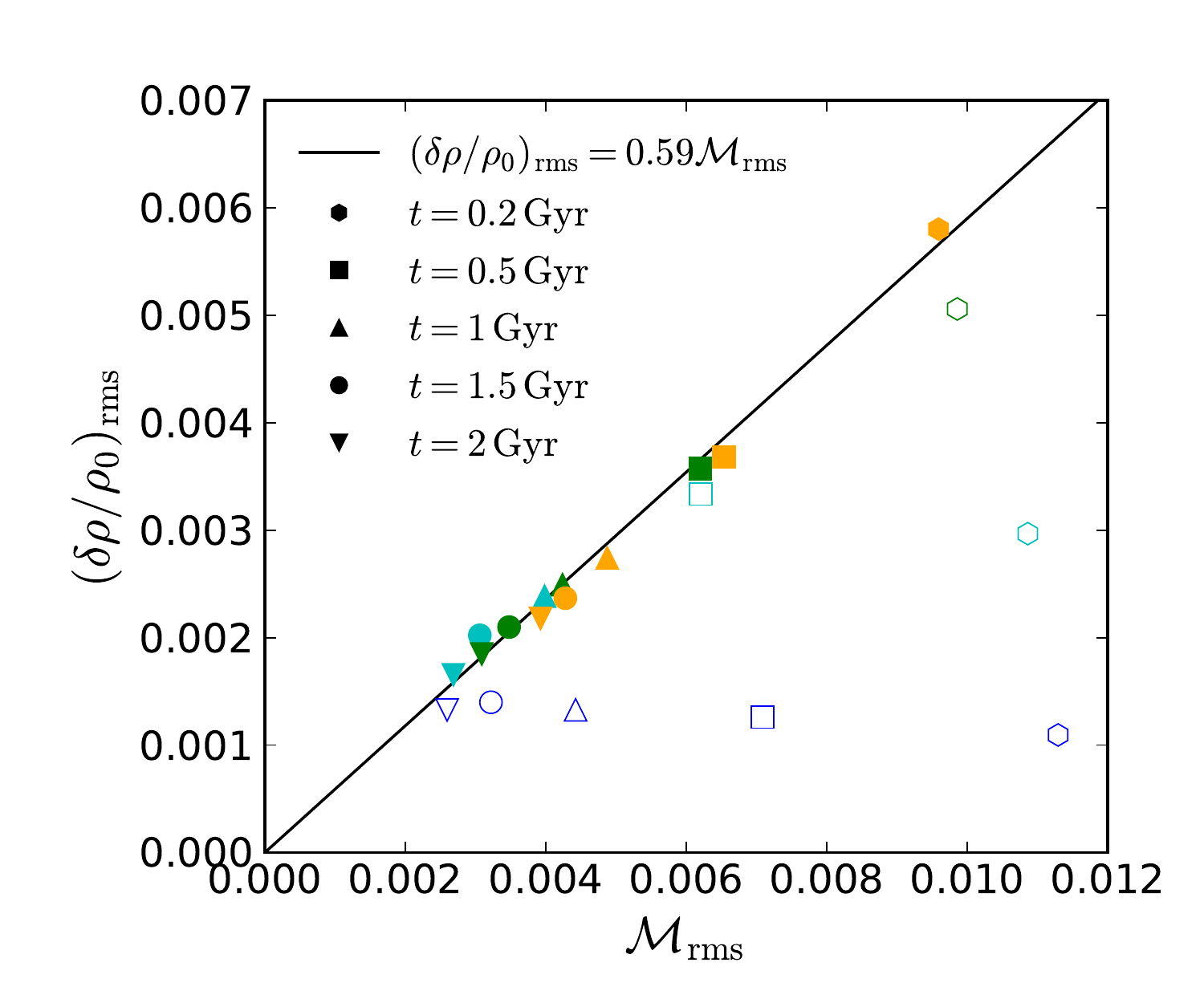}
\caption{Density fluctuation - Mach number relation in the energy equipartition
regime. The blue, cyan, green and orange colors indicate simulation runs with
$N_{\rm BV} = $ 0.7, 2.6, 5.3 and 10.6 Gyr$^{-1}$ respectively. Different markers show
snapshots at $t=$0.2, 0.5, 1, 1.5 and 2 Gyr. Snapshots that have obtained
saturated energy ratios (with $t > 2/ N_{\rm BV}$) are plotted in filled
markers, whereas those which have not are plotted in open markers.}
\label{fig:rhov_scatter}
\end{figure}

\begin{figure*}
\centering
\includegraphics[width=.9\textwidth,right]{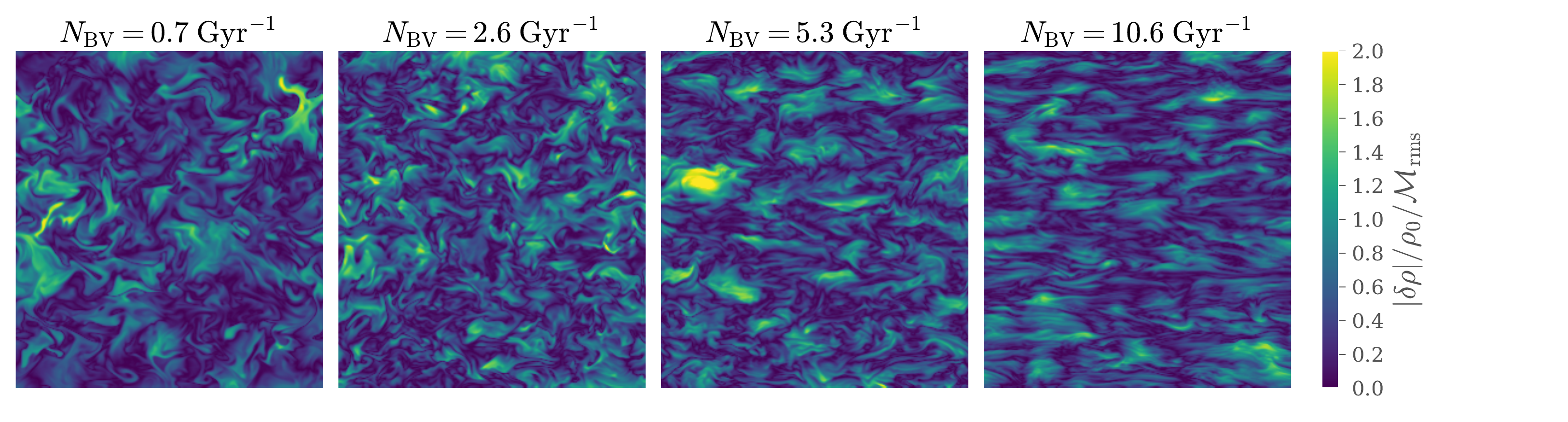}
\\
\includegraphics[width=.9\textwidth,right]{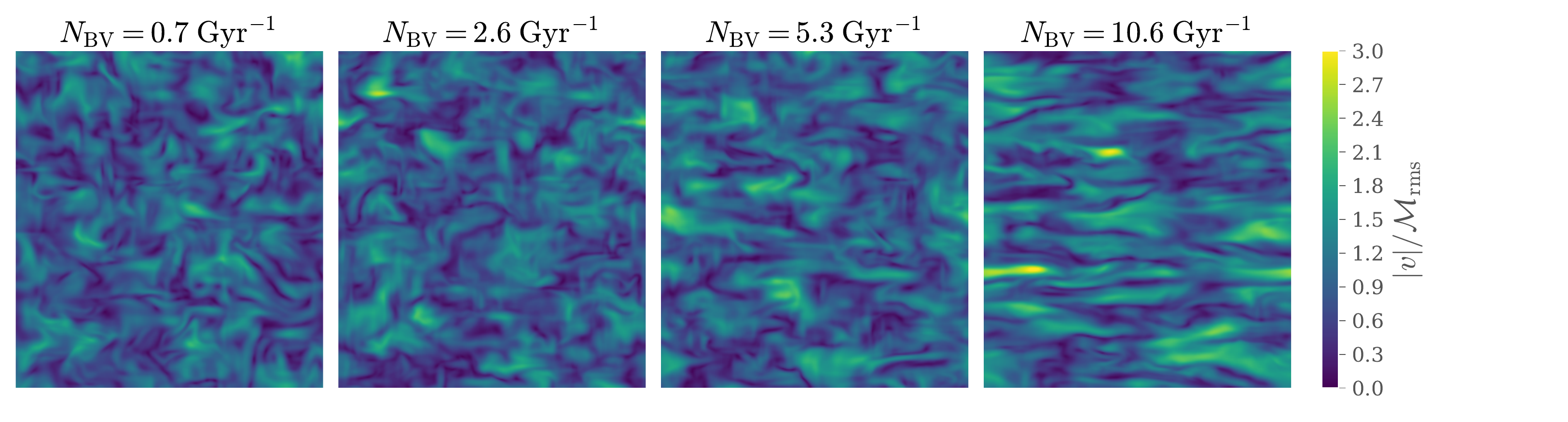}
\caption{Vertical slices of density fluctuation (upper panels) and velocity
magnitude (lower panel) at $t=$ 2 Gyr.}
\label{fig:v_2Gyr_xy}
\end{figure*}

Density fluctuations in the ICM are in general associated with ICM motions.
There have been attempts to infer the latter using X-ray observations of the
density fluctuations \citep{schuecker04, churazov12,
walker15, zhuravleva18}, which would rely on a correct theoretical
understanding of the relation between the two quantities.

In a homogeneous, isotropic turbulent medium, the amplitude of density
fluctuations depends predominantly on the r.m.s. turbulence
Mach number $\mathcal{M}_{\rm rms}$ and the associated compressibility of the
velocity field. For a qualitative understanding of this dependence, it is
most convenient to observe the equation of motion for the gas
Eq.\;\ref{eq:eom}. The divergence of Eq.\;\ref{eq:eom} leads to the
pressure Poisson equation in the form of \citep[e.g.][]{landau59} 
\eq{ \nabla^2 P = - \br{\nabla \vek{v}}^{T} : \nabla \vek{v}\,
}
when the motions are solenoidal i.e. $\nabla \cdot \vek{v} = 0$ and the 
medium is incompressible. The first order term in velocity on the l.h.s.
of Eq.\;\ref{eq:eom} does not enter due to the assumption of solenoidal
motions, and the lowest order dependence of gas pressure or density on velocity
is quadratic.
Thus, when the turbulence is driven with solenoidal modes, as the compressible modes do not develop in the low Mach
number limit $\mathcal{M}_{\rm rms} \ll 1$, the r.m.s. density fluctuation
$\ba{\delta \rho / \rho}_{\rm rms}$ scales as $\mathcal{M}_{\rm rms}^2$ at the lowest order. At
a finite Mach number, the compressible modes including sound waves are excited,
the first order term in velocity in the equation of motion starts to gain
dominance, leading to a linear increase of the density fluctuation with
$\mathcal{M}_{\rm rms}$, and thus a flattening in the density fluctuation -
turbulence Mach number scaling \citep[see also][]{pan10, mohapatra19}. This
trend continues to the supersonic regime, and the scaling approaches a logarithmic
form \citep{molina12} in the high Mach number limit where the velocity field is dominated by shock waves.

Even a moderate density stratification alters the relation completely. In a
density-stratified medium, the buoyancy-induced $\ba{\delta \rho / \rho}_{\rm
rms}$ scales linearly with $\mathcal{M}_{\rm rms}$, even in the low Mach
number limit (Eq.\;4 in \citealt{zhuravleva14}). Other effects such as perturbations of the gravitational
potential and thermal conduction \citep{churazov12,gaspari13,gaspari14} also
affect the relation (see discussion session).

Our simulations present a case study of the density fluctuation - velocity
connection in the low Mach number regime that is induced by buoyancy alone.
We note two facts that are often neglected in the literature. First, in a non-equilibrium system
like the ICM in a growing galaxy cluster, the coefficient in the linear relation 
between $\ba{\delta \rho / \rho}_{\rm
rms}$ and $\mathcal{M}_{\rm rms}$ is in general a time variable, and
it also depends on the density stratification and thus cluster radius, as shown by
Figs.\; \ref{fig:compare_power_evo_low_mach} and \ref{fig:power_evo} (note that
$e_{\rm pot} \propto \ba{\delta \rho / \rho}_{\rm
rms}^2$ and $e_{\rm kin} \propto \mathcal{M}_{\rm rms}^2$).
Only when the system is allowed to relax for a relatively long time ($t \gtrsim
1/N_{\rm BV}$) and also in the absence of significant changes of the gravitational potential, the coefficient saturates
to a nearly constant value presented in Fig.\;\ref{fig:rhov_scatter}.
Second, the linear relation holds only on a scale much larger than max($\ell_{\rm g}$,
$L_{\rm b}$). As shown by Fig.\;\ref{fig:v_2Gyr_xy}, there is no
local correspondence between the density fluctuation and the amplitude of
turbulence velocity. This can be understood by recalling the fact that the
former is a passive scalar tracer of the latter. The morphology of the
density field keeps a memory of the velocity field in the past.

\section{Discussions}
\label{sec:discussion}
\subsection{Global variations of gas density and gravitational
potential}
In our study, we have examined turbulence decay in a stable medium
where the background gravitational potential and gas density stay fixed
throughout the evolution. In reality, the merger events that inject turbulence into the ICM
would also trigger variations of the gas density and the gravitational
potential. These variations will couple to the turbulence
decay process and alter our results on turbulence energy evolution presented in
Sect.\;\ref{sec:energy_evo}, which we will discuss below. 

We consider here the turbulence evolution after a head-on major merger
event which, when present, dominates ICM turbulence injection \citep{paul11,
nelson12}. Especially, we focus on the evolution following the core collision
which marks the time of most intensive turbulence injection \citep{shi18}. As is
shown in \citet{shi18}, there is a fast decay phase lasting for about 1 Gyr
starting from the core collision, during which the gravitational potential
rapidly flattens as the dark matter halos of the merging objects pass through and move
away from each other, and the merged ICM rapidly expands.

In the presence of density stratification, a rapid flattening of gravitational
potential after a major merger event would channel away the gravitational potential energy associated with density fluctuations.
The shortage of potential energy would then forge more conversion of turbulence
kinetic energy into potential energy. Therefore, instead of reaching a balanced state
where both energies decay in power-law fashion as shown in
Figs.\;\ref{fig:compare_power_evo_low_mach},
\ref{fig:compare_power_evo_low_mach_NBVt_2panel} and \ref{fig:power_evo}, a fast
channel of turbulence kinetic energy loss would be formed in the system in
comparison to the regular turbulence cascade. The time scale for this fast
energy loss is characterized by that of the kinetic-potential energy conversion
i.e. the buoyancy time $1/N_{\rm BV}$. This mechanism may help the inner regions with stronger density stratification maintain a lower level of turbulence kinetic energy for $\sim$Gyr time before the final reversal of the radial
dependence when the residue trapped horizontal vortical modes gain dominance.
Considering that turbulence
injection by merger events is in general frequent in the ICM, and that each
injection is also extended in time \citep[e.g.][]{zhangcy16}, this final
reversal may never be reached. Thus, in this scenario, an increasing
turbulence amplitude with radius is the generic state of the ICM, in accordance
with the previous numerical results of ICM turbulence \citep{lau09, bat12, nelson14b, shi18}.

At the same time, a rapid expansion of the ICM following a major merger event
will directly cause adiabatic cooling of both the thermal and the kinetic
components of the ICM \citep{robertson12}, which also means a fast decay channel of the ICM
turbulence. In contrast to the mechanism caused by gravitational
potential flattening, this mechanism does not depend on ICM density
stratification. Nevertheless, it can introduce a radial dependence if the degree
of ICM density expansion is radial dependent. 

The above discussion has not considered the role of bulk motions in the
energy flow, which can be important when the gas cores of the progenitors are not destroyed
in the merger. It also neglects secondary effects such as the variation of the
Brunt-V\"ais\"al\"a frequency due to the flattening of gravitational potential
and the evolution of the entropy profile. Also, the relative importance of the
gravitational potential flattening effect and the ICM expansion effect to the
ICM turbulence evolution is not clear. All these would require further
systematic studies with more realistic numerical simulations. Nevertheless,
one hint is presented in the case study of \citealt{shi18} (see their
Fig.\;6). There, the time evolution of the ICM velocity fields after a major
merger is found to resemble that of the gravitational potential rather than the
gas density, suggesting a direct interaction between gravity and turbulence and
the dominance of the gravitational potential flattening effect.

\subsection{Thermal conduction}
Thermal conduction, which we neglect in this paper, can potentially influence 
the dynamics of the ICM and particularly the evolution of ICM turbulence. 
If thermal conduction is efficient, the
characteristic frequency of buoyant oscillations is not described by the
Brunt-V\"ais\"al\"a frequency as defined in Eq.\;\ref{eq:NBV} but with the entropy gradient in Eq.\;\ref{eq:NBV} replaced
by the temperature gradient \citep[e.g.][]{ruszkowski11}, reducing the
efficiency of turbulent kinetic energy - gravitational potential energy
conversion, and consequently reducing the buoyancy-induced density fluctuations
in the ICM \citep{gaspari14}.

Turbulence, on the other hand, plays an important role in determining the
potency of ICM thermal conduction by re-orienting and amplifying ICM magnetic
fields, with the latter suppressing ICM thermal conductivity compared to its
un-magnetized Spitzer value. Due to the complex interplay between ICM
thermal conduction, turbulence, magnetic fields and buoyancy, the efficiency of
thermal conduction in the ICM remains highly uncertain. However, recent studies
of both microscopic physics \citep[e.g.][]{komarov16, komarov18, robergclark18}
and macroscopic ICM structural stability \citep[e.g.][]{fang18} suggest that
the effective thermal conductivity in the ICM may be very small.

\section{Conclusions}
\label{sec:conclusion}
We have explored how the density stratification in the intracluster medium
affects the ICM turbulence time evolution, morphology, and the induced density
fluctuations with the aid of 3D hydrodynamic simulations. 
In particular, we examine if the Brunt-V\"ais\"al\"a buoyancy
frequency, which decreases by at least an order of magnitude from the 
central region of a galaxy cluster to its virial radius, would introduce a
radial dependence to the ICM turbulence evolution. The main results are as follows.\\

\begin{itemize}
\item \textit{The Froude number of ICM turbulence} \\
ICM turbulence is typically mildly-stratified with a Froude number $Fr \sim
\mathcal{O}(1)$. In the central region, the Froude number can reach as low as
$Fr \sim \mathcal{O}(0.1)$, which is still larger than the condition $Fr <
0.1$ commonly used to identify strongly-stratified turbulence. However, as we
have shown in this paper, the energy evolution, morphology and density
fluctuation of ICM turbulence with such a range of Froude number already show
a qualitative difference from those of homogeneous, isotropic turbulence. \\

\item \textit{Energy evolution of ICM turbulence}\\
The time evolution of turbulence energy in the density-stratified ICM
deviates from that in a homogeneous, isotropic medium. The buoyancy effect of a
density stratification lets turbulence generate density fluctuations, converting
part of its vertical kinetic energy to fluctuation potential energy and back on
a time scale of the buoyancy time $1/N_{\rm BV}$.
When the gravitational potential is kept constant, the kinetic and potential
energies approach a balanced state at long times ($t \gtrsim
1/N_{\rm BV}$), where both evolve in a
power-law fashion with a slope shallower than the $n\approx -1.5$ slope for
homogeneous isotropic turbulence.
We speculate that when the rapid flattening of gravitational potential
following a major merger event is taken into account, the conversion between
the two energies becomes one-way: turbulence kinetic energy converts quickly 
(with a time scale of $1/N_{\rm BV}$) into potential energy, which keeps
decreasing with the flattening of the gravitational potential.\\

\item \textit{Radial dependence of ICM turbulence}\\
In the more stratified central regions of the ICM, there is a
faster initial conversion from the injected turbulence kinetic energy to the
potential energy, resulting in an increasing turbulence amplitude with cluster
radius for $t \approx $ 0.2 Gyr. In our simulations where the 
gravitational potential is kept constant, this radial dependence will be
reversed later on due to a backflow of potential energy to vertical turbulence
kinetic energy, plus the trapping of more horizontal turbulence
kinetic energy as a consequence of suppression of vertical eddy turn-over in the
more density-stratified central regions. The inclusion of
gravitational potential flattening may help maintain the positive dependence of
turbulence amplitude on cluster radius for a longer time ($t \sim$ Gyr),
during which turbulence kinetic energy decays at a time scale characterized by
the buoyancy time $1/N_{\rm BV}$.
This is potentially the underlying physics for the faster turbulence decay in
the inner ICM regions discovered by \citet{shi18}, and that for the
increasing turbulence amplitude observed in cosmological numerical simulations.
\\

\item \textit{ICM turbulence morphology} \\
ICM velocity field is in general morphologically isotropic. However, 
in the case of turbulence in a cool core which has been quiescent for at
least $\sim$ Gyr, the Froude number reaches as low as
$Fr \sim 0.3$. In such a case, a clear density stratification
influenced morphology -- that of layered pancakes (vertically thin, horizontally
large vortical eddies) develops for the turbulence velocity field and the
density fluctuation it induces. The corresponding velocity anisotropy and the
layered density fluctuations are in principle detectable through X-ray observations.\\

\item \textit{Density fluctuation - Mach number relation} \\
A linear relation holds between the r.m.s. density fluctuation and the 
r.m.s. turbulence Mach number in a density-stratified medium even in the low
Mach number limit, but only when the quantities are averaged over a scale much
larger than both the turbulence eddy size and the buoyancy scale. We find that the
coefficient in the linear relation is in general a variable with radius
and time, which saturates to a constant only when the system is allowed to relax
for a relatively long time $t \gtrsim 1/N_{\rm BV}$, and when the gravitational potential is not
significantly changing with time.
\end{itemize}

\section*{Acknowledgements}
The software used in this work was in part developed by the DOE NNSA-ASC OASCR
Flash Center at the University of Chicago.

\bibliographystyle{mnras}

\end{document}